\documentclass[%
reprint,
aps,
pra
]{revtex4-2}

\usepackage[usenames]{color}
\usepackage{colortbl}
\usepackage{graphicx}% Include figure files
\usepackage{dcolumn}% Align table columns on decimal point
\usepackage{bm}% bold math
\usepackage{amsmath}
\usepackage{amssymb}
\usepackage[utf8]{inputenc}%Support cyrilic symbols
\usepackage{comment}
\usepackage{dsfont}
\usepackage{xcolor}

\begin{document}

\preprint{APS/123-QED}

\title{Thermodynamic coprocessor for linear operations with input-size-independent calculation time based on open quantum system 
}% Force line breaks with \\

\author{I. V. Vovchenko$^{1,3}$, A. A. Zyablovsky$^{1,2,3}$, A. A. Pukhov$^{1,2}$, E. S. Andrianov$^{1,2,3}$}
\affiliation{%
 $^1$Moscow Institute of Physics and Technology, 9 Institutskiy pereulok, Dolgoprudny 141700, Moscow region, Russia;
}%
\affiliation{
 $^2$Institute for Theoretical and Applied Electrodynamics, 13 Izhorskaya, Moscow 125412, Russia;
}
\affiliation{%
 $^3$Dukhov Research Institute of Automatics (VNIIA), 22 Sushchevskaya, Moscow 127055, Russia;
}%

%\affiliation{%
% $^4$Kotelnikov Institute of Radioengineering and Electronics, Mokhovaya 11-7, Moscow, 125009, Russia
%}%

\date{\today}% It is always \today, today,

\begin{abstract}
Linear operations, e.g., vector-matrix and vector-vector multiplications, are core operations of modern neural networks.
To diminish computational time, these operations are implemented by parallel computations using different coprocessors.
In this work we show that an open quantum system consisting of bosonic modes and interacting with bosonic reservoirs can be used as an analog thermodynamic coprocessor implementing multiple vector-matrix multiplications with stochastic matrices in parallel.
Input vectors are encoded in occupancies of reservoirs, and the output result is presented by stationary energy flows.
The operation takes time needed for the system's transition to a non-equilibrium stationary state independently on the number of the reservoirs, i.e., on the input vector dimension.
With technological limitations being considered, a device of $5\times5$ cm$^2$ area covered with the coprocessors can conduct of the order of $10^{11}$ operations per second per a mode of the OQS.
%The operation takes $10^{-9}\div 10^{-12}$ s to be implemented (depending on Q-factor).
The computations are accompanied by an entropy growth.
We construct a direct mapping between open quantum systems and electrical crossbar structures frequently used in analog vector-matrix multiplication, showing that dissipation rates multiplied by open quantum system's modes frequencies can be seen as conductivities, reservoirs' occupancies can be seen as potentials, and stationary energy flows can be seen as electric currents. 
\end{abstract}

\maketitle

%\tableofcontents

%\section{Introduction}
\textit{Introduction.}
Vector-matrix multiplication, particularly with stochastic matrices, is permanently exploited nowadays for natural language processing and steganography~\cite{almutiri2022markov,zekri2024large,dai2010text,sigaud2013markov}, decision making~\cite{bauerle2011markov,alagoz2010markov,an2008decision}, queuing theory~\cite{bocharov2011queueing}, %distributed control in multiagent systems~\cite{ren2008distributed,ren2007information}, 
solving optimization problems such as the traveling salesman problem~\cite{gunecs2024traveling,mo2011biogeography,martin1991large}, traffic optimization~\cite{salman2018alleviating,ahmad2019stochastic,xu2022approach,liu2020modeling}, and for different types of
forecasting, e.g., weather~\cite{yang2011first,eberle2022multivariate}, economy~\cite{phelan2022applications,kostoska2020structure,moosavi2017markovian,bauerle2011markov}, production~\cite{feldman2017revenue,dong2019pricing,fallahnezhad2020markov,ezugwu2017markov}, social opinion~\cite{jackson2008social,banisch2012agent,zhou2025novel}. 
%and combat actions~\cite{reese1971finite,washburn2009combat,biggs2023small}. 
In large-scale problems this operation is done by digital or analog coprocessors that implement it faster than a central processing unit (CPU) due to parallel computations~\cite{jawandhiya2018hardware,berggren2020roadmap,buber2018performance}.
Some of the used platforms overcome the von Neumann bottleneck by in-memory computations 
%(merging of memory and computational elements)
~\cite{wali2024two,seok2024beyond,bhaskaran2024phase,suri2020applications,liu2022memristor,de2021memristors} employing a crossbar structure (CS) for analog vector-matrix multiplication~\cite{xia2019memristive,cerofolini2009crossbar,li2021memristive,kim2012functional,ruhrmair2010applications,chen2003nanoscale,9932877,zhang2018neuromorphic}.
Particularly, the CSs are used in memristor-based neural networks (MNNs) %where their primary function is to implement vector-matrix multiplication
~\cite{adamatzky2013memristor,de2021memristors,suri2020applications,liu2022memristor} allowing MNNs to demonstrate low energy consumption and high suitability for neuromorphic computations \cite{wali2024two,seok2024beyond,bhaskaran2024phase,suri2020applications,de2021memristors,liu2022memristor,adamatzky2013memristor,stasenko2024astrocyte,shchanikov2023neuromorphic}.
%One possibility to realize iMC is to use crossbar structures that are employed in many modern implementations of in-memory computations and information storage~\cite{xia2019memristive,cerofolini2009crossbar,li2021memristive,kim2012functional,ruhrmair2010applications,chen2003nanoscale,9932877}.

A planar CS is a set of parallel conducting bars connected by memory (computing) elements with a similar set of perpendicularly oriented bars~\cite{li2021memristive,ruhrmair2010applications}.
In MNN, semiconducting memristors are used for computations~\cite{de2021memristors,seok2024beyond,wali2024two,mishchenko2022inverted,adamatzky2013memristor}.
%while MNN is realized as a matrix of memristors connected by a planar crossbar structure (further, crossbar structure) %that is used for analog processing of vector-matrix multiplication  
%\cite{de2021memristors,suri2020applications,liu2022memristor,adamatzky2013memristor}.
Usage of the memristors performing at quasi-particles~\cite{spagnolo2022experimental,zhang2020photon,cheng2022atomic,youngblood2023integrated,mao2019photonic,shrivastava2023fully} and transition to quasi-particle MNN (QPMNN) can increase speed and efficiency of the network~\cite{mao2019photonic,shrivastava2023fully,lian2022photonic,carroll2016photonic}.
For that, a proper CS should be designed to transmit the quasi-particles.
With miniaturization,
%The smaller this structure is, the more working nodes can be placed per unit area.
this technology will face quantum limit \cite{van2003moore,wu2013nanotechnology,randall2018digital}.
Thus, the design of a quantum CS, i.e., vector-matrix multiplayer, is a state-of-art {physical} problem.
%is more a state of art problem than an immediate necessity.
%but should be addressed preliminarily.

%In a CS particles transits between regions of space that have different matter and field composition.
The problem of the quantum CS design is very similar to the problems of quantum transport~\cite{yeyati2024photonic,kohler2005driven,nazarov2009quantum,haug2008quantum} and the physics of open quantum systems (OQSs)~\cite{breuer2002theory,vovchenko2024transient,binder2018thermodynamics,gemmer2009quantum}. %in the definition.
%A precise investigation of this can be done by means of the open quantum systems' physics
%non-equilibrium quantum thermodynamics 
%\cite{breuer2002theory,carmichael2009open,rivas2012open,vinjanampathy2016quantum,haug2008quantum,kosloff2013quantum,alicki2018introduction,gemmer2009quantum}.
In OQS' physics, the non-Hermitian dynamics of a quantum system, i.e., OQS, connected to reservoirs (environments, i.e., sources of thermal noise), including dynamics of energy, particle, and heat flows~\cite{breuer2002theory,levy2014local,trushechkin2016perturbative,vovcenko2023energy,vovchenko2024transient,potts2021thermodynamically,schaller2014open,pekola2021colloquium} is investigated~\cite{breuer2002theory,carmichael2009open,rivas2012open,vinjanampathy2016quantum,binder2018thermodynamics,gemmer2009quantum}.
Reservoirs are considered large enough compared to the OQS.
This allows to exclude their dynamics from the whole system's quantum dynamics by Born-Markov approximation~\cite{breuer2002theory,carmichael2009open}.
%It provides a well-developed toolkit to compute energy, particle, and heat flows~\cite{breuer2002theory,levy2014local,trushechkin2016perturbative,vovcenko2023energy,vovchenko2024transient,potts2021thermodynamically}.
%Thus, it suits the underlined problem.
%\cite{pascal2011circuit,thomas2019photonic,dey2023negative,pekola2021colloquium,yeyati2024photonic}.
After that, energy, particle, and heat flows between reservoirs connected via OQS can be calculated from the OQS's dynamics~\cite{potts2021thermodynamically,shishkov2019relaxation,vovchenko2025autonomous}.
During this dynamics, entropy is growing~\cite{vovchenko2025autonomous} and OQS reaches its non-equilibrium stationary state~\cite{binder2018thermodynamics}, and stationary energy, particle, and heat flows from reservoirs are established~\cite{vovcenko2023energy,vovchenko2024transient,potts2021thermodynamically}.

Recently, the transition of OQS to its stationary state was proposed to implement matrix inversion and to solve linear system of equations~\cite{aifer2024thermodynamic,melanson2025thermodynamic}.
Also, qubit-based OQS in Markovian limit was proposed as an adder~\cite{lipka2024thermodynamic,tiwari2025quantum}.
Such an involvement of the inherent thermodynamic physical features into the computational process is of interest {as investigation of new physical approaches and systems for implementation of computations has become an acute physical problem nowadays~\cite{aifer2025solving,whitelam2026generative,whitelam2026nonlinear,silva2026thermal,csaba2020coupled,todri2024computing,torrejon2017neuromorphic,wang2022optical,zuo2019all,berggren2020roadmap,seok2024beyond,bhaskaran2024phase,slinkov2025all}.}
The thermodynamic approach to computations shows great speed and low energy consumption~\cite{melanson2025thermodynamic}.
This is important in the modern trend of rapidly growing data centers' energy demand~\cite{IEA2025,IEA_Energy_and_AI}.

In this letter, we show that OQS consisting of bosonic modes can be used as a {thermodynamic} analog coprocessor implementing multiple vector-matrix multiplications with stochastic matrices in a parallel analog way.
Input vectors are encoded in reservoirs' occupancies, and output results are encoded in stationary energy flows.
This takes the time needed for the transition of the OQS to its non-equilibrium stationary state.
This time does not depend on the number of reservoirs and {afford $10\div1000$~TOps/s (terraoperations per second) on the $5\times 5$~cm$^2$ area per a mode of the OQS.
However, nowadays technological opportunities in temperature manipulation reduce this computational rate to $100$~GOps/s per.}
The consideration {of the OQS's dynamics} is done by means of the global approach to dissipation~\cite{hofer2017markovian,potts2021thermodynamically,vovchenko2021model}, i.e., the second law of thermodynamics is fulfilled and computations are accompanied with entropy growth.
%The usage of thermal noise for computations and natural entropy growth accompanying the OQS's evolution makes the computations more fault-tolerant.
We develop an electrical analogy for the OQS, showing that it can be represented as a planar CS circuit, i.e., dissipation rates multiplied by OQS's modes frequencies can be seen as conductivities, reservoirs' occupancies as potentials, and stationary energy flows as electric currents.
%interacting with several reservoirs that have different temperatures (and chemical potentials) considering the Markovian limit of OQSs' dynamics. 
%This relates the research with the fields of thermal linear algebra, thermodynamic computing and thermal neural networks~\cite{aifer2024thermodynamic,melanson2025thermodynamic,lipka2024thermodynamic}.

%limit that mimics stationary energy flows through the open system with quadratic interaction via electric currents.
%This open system is connected to several reservoirs with different temperatures (and chemical potentials).
%We present an electrical circuit that reproduces stationary energy flows in the problem via electric currents.
%We show that this electrical circuit can be presented in the form of cross-bar scheme.
%We show that dissipative rates at power minus one can be seen as resistances and reservoirs' occupancies and temperatures can be seen as potentials.
%We apply developed electrical analogy to describe a circuit that exhibits diode-like behaviour, being able to block inverse currents.
%Also we show that it is possible to block energy flows at several frequencies simultaneously with help of additional reservoirs.

%\section{Implementation of linear algebra operations in OQS}

\begin{figure}
\center{\includegraphics[width=\linewidth]{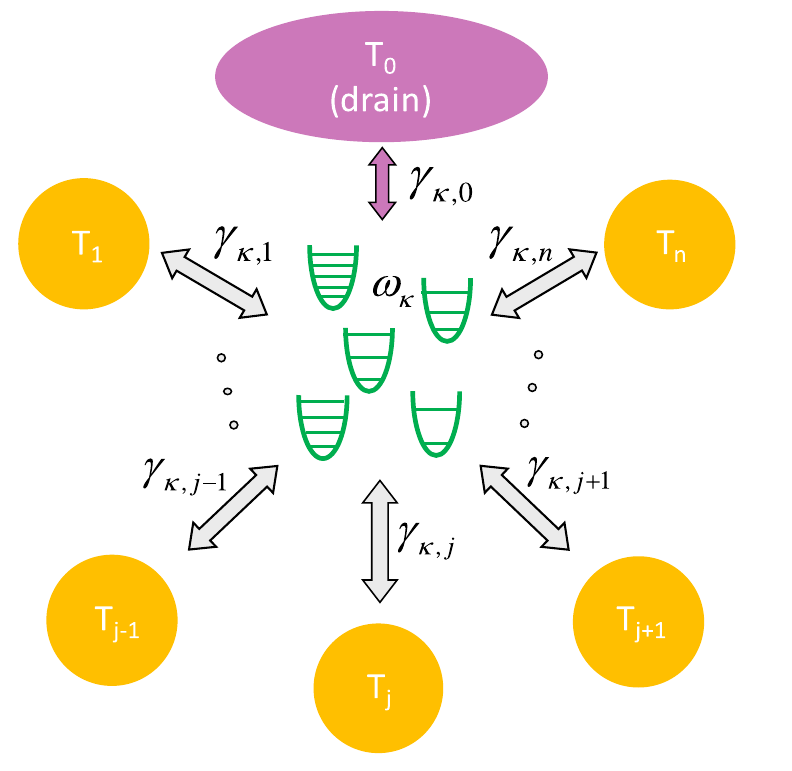}}
\caption{Schematic representation of the considering OQS.}
\label{Open_syst}
\end{figure}

\textit{The model.} We consider an OQS consisting of $K$ bosonic modes, see Fig.~\ref{Open_syst}, that is a common model for photons~\cite{hartmann2006strongly,hartmann2016quantum,wang2024realization,roushan2017chiral,roushan2017spectroscopic}, phonons~\cite{minarik2013hamiltonian,zaitsev2008introduction,scherer1984theory}, and magnons~\cite{yuan2017magnon,Kittel1991quantum,mcclarty2022topological}.
The Hamiltonian of this OQS reads \cite{hartmann2006strongly,hartmann2016quantum,asadian2013heat}
$\hat H_S=\sum_{\kappa=1}^K \omega_\kappa \hat a^\dag_\kappa \hat a_\kappa.$
Here, $\omega_\kappa$ are frequencies of the modes, $\hat a_\kappa$ are lowering operators of the modes, $[\hat a_v,\hat a_w]=0$, $[\hat a^\dag_v,\hat a_w]=-\delta_{vw}$.
The number $K$ is arbitrary, but further we show that it defines the number of possible parallel vector-matrix multiplication operations in the OQS. 
%$\hat H_S=\sum_{v,w} \varepsilon_{vw} \hat c^\dag_v \hat c_w=\sum_\kappa \omega_\kappa \hat a^\dag_\kappa \hat a_\kappa.$
%Here $\varepsilon_{vw}=\varepsilon_{wv}\in \mathds{R},\ \hat c_v$ are lowering operators of oscillators, and $\hat a_\kappa$ are lowering operators of $\hat{H}_S$ modes, $[\hat c_v,\hat c_w]=[\hat a_v,\hat a_w]=0$, $[\hat c^\dag_v,\hat c_w]=[\hat a^\dag_v,\hat a_w]=-\delta_{vw}$.
%Operators $\hat{c}_v$ and $\hat{a}_\kappa$ are linked by unitary transformation~\cite{bellman1997introduction,bogolubov2010introduction,colpa1978diagonalization}.

%There is known electrical analogy in quantum transport usually addressed as circuit approach \cite{}.
%It is usually used

There are many approaches to describe the dynamics of the OQS interacting with a number of reservoirs~\cite{potts2024quantum,binder2018thermodynamics,breuer2002theory}.
In the Markovian limit, local and global approaches are usually used~\cite{potts2021thermodynamically,vovchenko2021model,vovcenko2021dephasing,vovcenko2023energy,cattaneo2019local}.
However, under the local approach, second law of thermodynamics can be violated~\cite{levy2014local,vovcenko2023energy}, while under the global approach it is always fulfilled~\cite{spohn1978entropy}.
With the aim to ensure the fulfillment of the second law of thermodynamics, we use the global approach.

We suppose that the OQS interacts with $n+1$ reservoirs at temperatures $T_j$, where $j$ goes from $0$ to $n$.
Further, we assume that $T_0\ll T_{j\ne 0}$ and use this cold reservoir as a drain.
The energy flow from the $j$-th reservoir to the OQS in the global approach 
% (Davies master equation approach) 
~\cite{breuer2002theory,gorini1976completely,davies1974markovian} equals~\cite{vovchenko2024transient}
\begin{gather}\label{J_el}
{J}_j \hspace{-2pt}
=\hspace{-3pt}
\sum_{\kappa=1}^K\hspace{-2pt} {J}_{\kappa,j}\hspace{-2pt}
=\hspace{-2pt}
\sum_{\kappa=1}^K \hspace{-1pt}\omega_\kappa \gamma_{\kappa,j}
(n_j(\omega_\kappa, T_j)\hspace{-2pt} - \hspace{-2pt}\tilde{n}(\omega_\kappa,T_{0,\cdots, n})).
\end{gather}
Here $\gamma_{\kappa,j}$ is dissipation rate of the OQS's $\kappa$ mode (with frequency $\omega_\kappa$) to the $j$-th reservoir, $\tilde{n}(\omega_\kappa,T_{0,\cdots, n})=\sum_{j=0}^n p_{\kappa,j} n_j(\omega_\kappa, T_j)$
is weighted occupancy among reservoirs at frequency $\omega_\kappa$, i.e., $p_{\kappa,j}=\gamma_{\kappa,j}/\sum_{m=0}^n \gamma_{\kappa,m}$, $n_j(\omega_\kappa,T_j)=(\exp (\omega_\kappa/T_j)-1)^{-1}$ is occupancy of the $j$-th reservoir at frequency $\omega_\kappa$, ${J}_{\kappa,j}$ denotes energy flow through the frequency $\omega_\kappa$.
{Note that the $\omega_\kappa$ used here from~\cite{vovchenko2024transient} can be seen as eigenmodes' frequencies of an diaganolized quadratic Hamiltonian that takes the coupling between modes into the account~\cite{bellman1997introduction,bogolubov2010introduction}.
Here and further, we address these eigenmodes as modes.
This approach is reliable when nonlinearities in the OQS or its modes' amplitudes are small~\cite{klimov2000method,gong2009effective,leonski2004kerr}.
Then the contribution of nonlinearities can also be taken into account by a diagonalizable quadratic Hamiltonian~\cite{klimov2000method,zaitsev2008introduction,Landau_StatPhys2}.
}
%The value $\tilde{n}(\omega_\kappa,T_{0,\cdots, n})$ is weighted occupancy among reservoirs at frequency $\omega_\kappa$.

{Eq.~(\ref{J_el}) allows for a transparent physical interpretation.
A quantum with $\omega_\kappa$ frequency from the OQS can transfer to one reservoir only, as is the Markovian description the OQS and the reservoirs are dissentagled~\cite{shishkov2019relaxation,carmichael2009open,breuer2002theory}.
This is a probabilistic process, with probability proportional to the OQS dissipation rate to the particular reservoir.
Hence, the OQS's quanta exhibit $p_{\kappa,j}$ average rate of transfers to the $j$-th reservoir.
In this relaxation process, the occupancies of the OQS's modes tend to the values of weighted occupancies of the reservoirs, i.e., to their stationary values.
These stationary occupancies, as we show further, regulate the stationary energy flows between the reservoirs similarly to the way electric potentials in circuits regulate the electric currents in nodes connected to different sources.}

{This dependence of the energy flows in Eq.~(\ref{J_el}) on weighted occupancies $\tilde{n}(\omega_\kappa,T_{0,\cdots, n})$ among the reservoirs can be used to implement linear algebra operations as discussed below.}

\textit{The scalar product.}
Eq. (\ref{J_el}) can be rewritten as follows
\begin{gather}\label{J_el1}
    {J}_j=\sum_{\kappa=1}^K {J}_{\kappa,j}=\\ \nonumber
=\sum_{\kappa=1}^K \omega_\kappa \sum\limits_{q=0}^n \frac{\gamma_{\kappa,j} \gamma_{\kappa,q} }{\sum\limits_{m=0}^n\gamma_{\kappa,m}  }(n_j(\omega_\kappa, T_j) - n_q(\omega_\kappa,T_q)).
\end{gather}

We consider $T_0\ll T_{j \ne 0}$ such that $n_0(\omega_\kappa,T_0)\ll n_{j\ne 0}(\omega_\kappa,T_{j})$.
Then,
\begin{gather}\label{J_el2}
    {J}_0\approx -\sum_{\kappa=1}^K \omega_\kappa  \gamma_{\kappa,0} \sum\limits_{q=1}^n p_{\kappa,q} n_q(\omega_\kappa,T_q)=\\ \nonumber
    =-\sum_{\kappa=1}^K \omega_\kappa  \gamma_{0,\kappa}  (\vec{p}_{\kappa},\vec{n}(\omega_\kappa,\vec{T})),
\end{gather}
Here $\vec{T}=(T_1,\ldots,T_n)^T$ and
\begin{gather}
    \vec{p}_\kappa=
    \left[
    \begin{array}{cc}
         p_{\kappa,1}\\
         \vdots\\
         p_{\kappa,n}
    \end{array}
    \right],\ \ \ 
    \vec{n}(\omega_\kappa,\vec{T})=
    \left[
    \begin{array}{cc}
         n_1(\omega_\kappa,T_1)\\
         \vdots\\
         n_n(\omega_\kappa,T_n)
    \end{array}
    \right].    
\end{gather}
%Here $\vec{n}_1(\omega_\kappa,\vec{T})=(n_2(\omega_\kappa,T_2),\ldots,n_n(\omega_\kappa,T_n))^T$, $\vec{p}_{1,\kappa}=(p_{2,\kappa},\ldots,p_{n,\kappa})^T$.

It is seen that energy flow through the $\omega_\kappa$ mode is proportional to the scalar product of $\vec{p}_{\kappa}$ and $\vec{n}(\omega_\kappa,\vec{T})$ vectors.
This can be used to implement scalar product of positive vectors.
Indeed, suppose that we want to find $(\vec{a},\vec{b})$, where $\vec{a},\vec{b}\in \mathds{R}^{n}\ge 0$ (this denotes that elements of both vectors are non-negative).
Vector $\vec{a}$ is considered to be normalized. 
For that
\vspace{3pt}

(i) one needs to connect $n+1$ reservoirs through a set of bosonic modes, and one of the reservoirs should be cooled down $T_0\ll T_{j\ne 0}$ to fulfill the condition $n_0(\omega_\kappa,T_0)\ll n_{j\ne 0}(\omega_\kappa,T_j)$ for some modes $\omega_\kappa$ of the OQS (the drain reservoir);
\vspace{3pt}

(ii) we need to choose a mode from the modes mentioned above to conduct the calculations, let it be the $\omega_1$ mode;
\vspace{3pt}

(iii) we need to manipulate dissipation rates at frequency $\omega_1$ setting $\gamma_{0,1}\ll \sum_{j=1}^n \gamma_{j,1}$ to make $\vec{p}_1$ a normalized vector.
\vspace{3pt}

(iv) we need to manage dissipation rates at frequency $\omega_1$ and temperatures of other reservoirs to satisfy the relations 
\begin{gather}
\vec{p}_{1}=\vec{a}, \quad \vec{n}(\omega_1,\vec{T})=\vec{b};
\end{gather}
%\vspace{3pt}

(v) we need to analyze the spectrum of the energy outcome from the OQS to the drain reservoir and apply the equation $(\vec{a},\vec{b}) =-{J}_{1,0}/\omega_1 \gamma_{1,0}$ that follows from Eq.~(\ref{J_el2}).
\vspace{3pt}

For this procedure to be implemented, one needs to set $\vec{b}$ by managing reservoirs' temperatures.
This can always be done as $\partial n(\omega,T)/\partial T >0$.
Indeed, from $b_j=1/({\rm exp}(\omega_1/T_j)-1)$ we get $T_j=\omega_1/\ln{(1+1/b_j)}$.
Management of the dissipation rates can be done in many ways, particularly by managing the coupling between the OQS's modes and the modes of the reservoirs~\cite{poyatos1996quantum,myatt2000decoherence,mendoncca2020reservoir,mottonen2010probe,muschik2012robust,krasnok2015antenna,krasnok2016demonstration,chauhan2017motion,zyablovsky2017approach,nefedkin2018mode}.

Note that the mode of the OQS in this setting acts as an adder.
It sums its entries (reservoirs' occupancies) with given weights (dissipative rates) and provides the output signal (energy flow) that is proportional to the weighted sum of the entries.
One can restrict the values of the entries at a moderate level of the OQS modes' amplitudes to process normalized data.
Then the contribution of OQS's nonlinearities would be taken into account by the diagonalized OQS's Hamiltonian, as it is discussed above.

In parallel to the described procedure, different scalar products are calculated at other frequencies $\omega_\kappa,\ \kappa\ne 1$.
Indeed, as Eq.~(\ref{J_el2}) states, the energy flows at other frequencies of OQS are also proportional to some scalar products.
Occupancies of reservoirs at these frequencies are functionally dependent on occupancies at $\omega_1$.
Hence, we get additional $K-1$ scalar products of vectors that are functionally dependent on $\vec{b}$ with different vectors defined by dissipation rates of OQS at the frequencies $\omega_\kappa,\ \kappa\ne 1$.
This might be useful for feature extraction procedures in neural networks and object recognition~\cite{guyon2008feature,mutlag2020feature,li2018learning}.

For that, the frequency distance between the modes of the OQS should be sufficient to neglect the effect of non-resonant stationary energy transport, which is revealed in non-Markovian models of the OQS dynamics~\cite{haug2008quantum,schaller2014open,pekola2021colloquium}.
This effect is relevant when the frequency distance between OQS's modes is of the order of the dissipation rates.

%with the described procedure the scalar products on others $\omega_\kappa$ frequencies are calculated.
%Despite the fact that dissipative rates on different frequencies can be manipulated independently, vectors $\vec{n}_1(\omega_\kappa,\vec{T})$ are functionally dependent on the $\vec{n}_1(\omega_1,\vec{T})$ vector.
%Thus, the result of this parallel computation is not totally independent on the first one, but it might be of some interest in several applications~\cite{}.

\textit{The vector-matrix multiplication.}
Being able to implement the scalar product of a normalized positive vector with another positive vector, one can implement multiplication of an arbitrary stochastic matrix to a positive vector.
%as this operation implies only calculation of scalar products.
For that, we need to implement the scalar product several times.
This can be done by employing an array of considered systems or by employing other modes of the OQS.

The first option is trivial.
%This can be implemented, for example, using degenerate modes.
If one has $m$ copies of the system, then
\begin{gather}\label{vec_mat_mul}
\vec{{J}}_{1,0}=P_1\vec{n}(\omega_1,\vec{T}),\ \text{where}\\ \nonumber
    \vec{J}_{1,0} = \left[
    \begin{array}{cc}
    {J}_{1,0}^{(1)}\\
         \vdots\\
    {J}_{1,0}^{(m)}
    \end{array}
    \right],\quad P_1 = 
    \left[
    \begin{array}{ccc}
        p_{1,1}^{(1)} & \cdots  & p_{1,n}^{(1)}\\
        \vdots & \ddots  & \vdots \\
        p_{1,1}^{(m)} & \cdots & p_{1,n}^{(m)}
    \end{array}
    \right]. 
\end{gather}
Here ${J}_{1,0}^{(k)}$ is energy flow through the $\omega_1$ mode of the $k$-th copy of the OQS, and $p_{1,1}^{(k)}$ is $p_{1,1}$ in the $k$-th copy of the OQS.
Every row of the matrix $P_1$ is a normalized positive vector.
Hence, if $n=m$, then $P_1$ is a stochastic-row matrix, i.e., transition matrix of a Markov chain~\cite{bocharov2011queueing}.

Note that in parallel with this procedure, vector-matrix multiplication happens at other modes of the OQS.
This is similar to the way the parallel scalar product is implemented by different $\omega_\kappa$ described above.
%As before, dissipation rates on different frequencies can be manipulated independently.
%The matrices of normalized dissipation rate at different $\omega_\kappa$ are of different sizes, as the number of matrix rows depends on the mode's degeneracy $m_\kappa$.
Input vectors $\vec{n}(\omega_\kappa,\vec{T})$ at different modes are functionally dependent on the first $\vec{n}(\omega_1,\vec{T})$ input vector.
Thus, the results of these parallel computations are not totally independent but might be useful for feature extraction and object recognition~\cite{guyon2008feature,mutlag2020feature,li2018learning}.

To implement the second option
the input vectors $\vec{n}(\omega_\kappa,\vec{T})$ should be approximately the same at the used OQS's modes.
Thus, we need to use degenerate modes of the OQS or modes with close frequencies.
If one handles independent management of the dissipation rates of OQS's modes with different frequencies, this second option can be realized multiple times in the frequency domain of the OQS with several well-separated-by-frequency-gaps groups of the modes.
Along with the copying of the system, this can boost the total rate of computations per unit area.
%Let us suppose that the modes of OQS are split into $M$ groups with close frequencies ($M\le K$), i.e. inside the group $n_{j\ne 0} (\omega_\kappa,\vec{T})\approx n_{j\ne 0} (\omega_1,\vec{T})$, and groups are well separated one from the other in frequency domaine.
%{\color{black}We} denote by $m_1$ the number of OQS's modes that are close to $\omega_1$ such that $n_{j\ne 0} (\omega_\kappa,\vec{T})\approx n_{j\ne 0} (\omega_1,\vec{T})$.
%Applying the procedure described above to all these modes, we get 

The ability to manage dissipation rates can also be used to implement learning process in this system.
Indeed, with a feedback loop being designed, comparison of the computation result with the target result can be utilized for the dissipation rates' managing~\cite{doyle2013feedback,li2018learning}. 

%Discussed here procedures of scalar product and vector to matrix multiplication implementation requires some time to be fulfilled.
\textit{The device.}
In this section we discuss possible realization of the proposed computational scheme.
The important properties of a computational device are number of operations done per second and fault tolerance.
The presented scheme uses stationary energy flows from noisy environments for computations.
%Setting up dissipation rates and reservoirs' temperatures is an engineering problem.
%This time might be minimized by proper realization of the device.
The establishment of the flows requires time that is of the order of the OQS dissipation time and does not depend on the initial state of the OQS and the number of the reservoirs~\cite{vovchenko2024transient,vovchenko2021model}, if long-living modes (for example, subradiant modes) are avoided in the evolution of the OQS~\cite{willingham2011energy,jenkins2017many,zhou2011tunable,meng2023strong}.
%For $\omega/T\sim 1,$ and $T\sim500$ K, we have $\lambda\sim 10^{-4}$ m.
If, e.g., we consider an ensemble of resonators {(or one resonator, see Fig.~\ref{Fig_the_device})} in the GHz range as the OQS, interacting with the free space electromagnetic modes as the drain reservoir, then for the wavelength of the OQS mode $\sim 10^{-3}$ m and Q-factor (quality factor) $\sim 10^{2}\div 10^4$ that can be achieved simultaneously in the experiments with {high-resistivity silicon, multiple-ring} and photonic crystal resonators~\cite{otter2014100,krupka2016high,temelkuran2001quasimetallic,hsu2005w}, we can estimate the {computational time} as $10^{-8}\div 10^{-10}$ s.
This time is required to realize the above-discussed vector-matrix multiplication.
Notably, the lower the Q-factor is, the faster computations are done.
{(All the mentioned systems are suitable candidates for the role of the OQS and can be described using the same formalism, thus, further, we suppose that the OQS is just a resonator of millimeter length).}

%~\cite{young2013generating,lecamp2005energy,gregersen2010numerical,liu2022photonic,reitzenstein2007alas},
%However, if subradiant (dark) modes are present in OQS, they should be avoided in the computations as they can exhibit sufficiently larger dissipation time~\cite{willingham2011energy,jenkins2017many,zhou2011tunable,meng2023strong}.
%This boundary can not be lowered without tuning of the OQS's initial state, that we avoid.
%The results of the calculation does not depend on the initial state of the OQS, as stationary energy flows are analyzed.
%Thus, the time required for the considered vector-matrix multiplication is at least of the order of OQS's dissipation time, i.e. nanoseconds~\cite{}.
%Hence, it is important to avoid usage of subradiant modes in computations, as they can exhibit sufficiently large dissipation time~\cite{}. 

{
To implement the reservoirs of the OQS, one can use locally heated waveguides { made either from silicon nitride, silica, lithium niobate, or polymers}~\cite{yong2022power,nejadriahi2020thermo,van2010integrated,zhong2021fast,prencipe2023electro,xie2023polymer}, { see Fig.~\ref{Fig_the_device} for the schematic representation of the device.}
The heating can be provided by heaters that are of $10^{-5}$ m size~\cite{van2010integrated,zhong2021fast,xie2023polymer}.
The dissipation rates then are proportional to the square of evanescent coupling between the waveguides' modes and the modes of the OQS~\cite{vovchenko2024transient,yeh2008essence}.
This coupling can be controlled in multiple ways~\cite{poyatos1996quantum,myatt2000decoherence,mendoncca2020reservoir,mottonen2010probe,muschik2012robust,krasnok2015antenna,krasnok2016demonstration,chauhan2017motion,zyablovsky2017approach,nefedkin2018mode,yeh2008essence,huang1994coupled,haus2002coupled,hwang2022electro,li2023electro,witmer2017high} (for example, via management of the distance between the OQS and the waveguides).
Then, one can expect the $n\sim 100$ on the $10^{-3}$ m size of the OQS mode's wavelength and the computational rate of about $10\div 1000$~GOps/s per a mode of the OQS.

On the approximate CPU area of $5\times 5$~cm$^2$, it is possible to place of the order of $1000$ of the considered computing systems.
Hence, the possible computational rate of such a device per a mode of the OQS can be estimated as $10\div 1000$~TOps/s.

That large computational rate is a fundamental limitation of the computational rate per the limited area.
To reach it, the whole computational cycle should be done on the timescale of the OQS dissipation time.
Here, the technological limitations impose stricter restrictions.
The computational rate of the coprocessor in reality will be limited by the time of thermal response that is usually on the order of microseconds in the experiments~\cite{van2010integrated,zhong2021fast}.
Hence, the nowadays realistic value of the coprocessor computational rate should be estimated as $100$~GOps/s per the OQS's mode.

\begin{figure}
    \centering
    \includegraphics[width=0.8\linewidth]{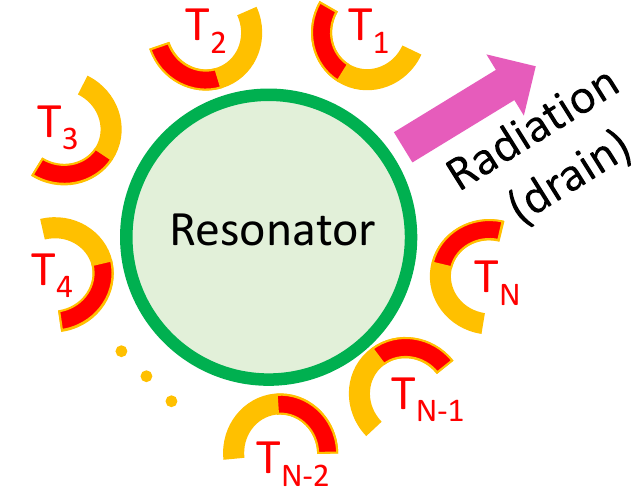}
    \caption{ Schematic representation of the proposed realization of the OQS and the reservoirs from Fig.~\ref{Open_syst}. 
    The open quantum system is represented by the modes of the microring resonator (green circle), the reservoirs are implemented by the waveguides (yellow semi-circles) heated by the heaters (red quarter circles).
    The drain reservoir is radiation of the OQS.
    }
    \label{Fig_the_device}
\end{figure}

This computational rate is comparable with the computational rate of some modern GPU solutions used in industry~\cite{reuther2020survey}.
Furthermore, it is of great interest in the context of a full-thermodynamic computer development~\cite{aifer2024thermodynamic,melanson2025thermodynamic,lipka2024thermodynamic,tiwari2025quantum,rolandi2026energy,whitelam2026generative,whitelam2026nonlinear} and extra computations in microelectronic systems conducted by spurious heat~\cite{silva2026thermal}.
Also, the development of faster heaters along with the design of the scheme allowing for more guided and direct heating effect can increase the computational rate of the device. 

Additionally, one can try managing dissipation rates of the OQS modes having different frequencies in parallel.
As it is discussed above, this can increase the computational rate in the system too.
For that, a proper choice of the waveguides and the medium that connect them to the OQS should be done to match the needed integral of overlapping between the electromagnetic fields of the OQS and the reservoirs at chosen frequencies~\cite{yeh2008essence,huang1994coupled,haus2002coupled}.

%Beyond that, one can design a bosonic OQS as metastructure to posses the needed mode composition and needed coupling to environments using artificial intelligence~\cite{}.
%Similarly, the conditions needed to conduct the vector-matrix multiplication can be added to the model~\cite{silva2026thermal}.
}

%The energy consumption of a modern thermo-optical element is of the order of $1$ mW~\cite{}.
%Hence, one elementary operation requires of the order of $10^{-9}$ J.
%Bad result.

%The presented scheme is an analog scheme, i.e. the result of the computations continuously depends on the problem parameters and input~\cite{care2010technology,maclennan2018analog}.
Note that the stationary energy flows in OQS in fact are mean stationary energy flows, as the fluctuations of temperatures are already averaged when Born-Markov approximation is valid~\cite{carmichael2009open,haake1973statistical,breuer2002theory}.
This delivers reliable results due to the fact that reservoirs' correlation functions decay faster than the OQS dissipates~\cite{carmichael2009open,haake1973statistical,breuer2002theory,haake1969non,vovchenko2024transient,cattaneo2019local}.
%Hence, environmental fluctuations do not contribute to the result of computations.
Also, the OQS's evolution to the stationary state is accompanied by entropy growth~\cite{breuer2002theory,Landau_StatPhys,spohn1978entropy}.
However, the OQS's entropy can increase or decrease in this process.
Nevertheless, this does not affect the result of the calculations.
Mentioned effects mask the contribution of random errors and increase the fault tolerance of the scheme.

Also, we should mention that the influence of the non-Markovian non-resonant stationary energy transport discussed above is negligible in the considering scheme.
Indeed, the dissipation rate of the OQS is inversely proportional to its dissipation time and defines the spectral width of the non-Markovian non-resonant stationary energy transport~\cite{haug2008quantum,schaller2014open}.
Thus, in the considering system, this spectral width is of the order of $10^{8}\div 10^{10}$ s$^{-1}$, while the frequency step between OQS's modes is of the order of $10^{11}$ s$^{-1}$.
Hence, the non-Markovian non-resonant stationary energy transport is irrelevant in the considering setup.

\begin{figure}
\center{\includegraphics[width=0.7\linewidth]{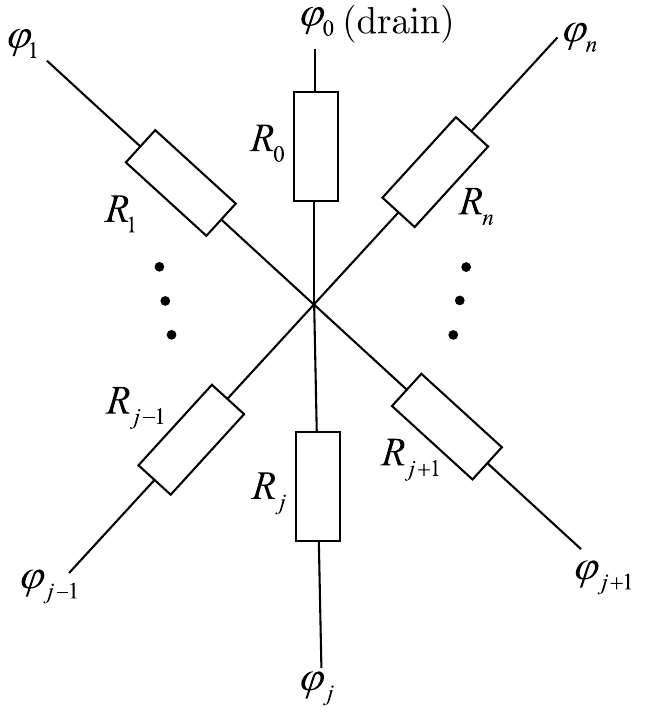}}
\caption{Electrical circuit supporting currents equivalent to the energy flow ${J}_{\kappa,j}(\vec{T})$ in Eq.~(\ref{J_el}) for a particular $\kappa$, {i.e., the energy flows through a frequency of the OQS.}}
\label{Scheme}
\end{figure}

The discussed realization of the thermodynamic coprocessor has restricted opportunities for the dissipation rates manipulation.
Thus, it is a worthwhile device when a multiplication of a particularly needed, once-established matrix by an arbitrary normalized positive vector should be done.
For instance, the considered realization of the coprocessor is suitable for performance in an already learned neural network.

\textit{Electrical analogy for the OQS.} 
As CS are usually used in microelectronics, it will be useful to have an electrical analogy of the process considered above to visualize the achieved results. 

{
As we have stated above, the stationary occupancies of the modes in the OQS act like potentials in electric circuits.
To show that we consider the electric circuit in Fig.~\ref{Scheme}.
}
%Here we develop it by mapping the OQS to a CS.
%We show that the energy flow in Eq.~(\ref{J_el}) can be associated with electric current in the electric circuit presented in Fig.~\ref{Scheme}.
In this circuit $n$ wires having resistances $R_j$ and potentials $\varphi_j$ applied to their ends are connected by a central node.
The potential formed in the central node {is denoted as} $\tilde{\varphi}$.
The electric currents $I_j$ by Ohm's law equal~\cite{alexander2007fundamentals} $\varphi_j - \tilde\varphi= I_j R_j$.
By Kirchhoff's first law, the sum of incoming electric currents in the node should be zero in stationary regime $\sum_j I_j=0$~\cite{alexander2007fundamentals}.
Hence, $\sum_j (\varphi_j - \tilde\varphi)/R_j=0$ and
\begin{equation}\label{phi0}
    \tilde\varphi=\frac{\sum_j \varphi_j /R_j}{\sum_j {1}/{R_j}}=\sum_j p_j \varphi_j.
\end{equation}
Here, $p_j=(1/R_j)/\sum_j (1/R_j)$.
Thus, $\tilde{\varphi}$ equals the weighted potential of wires, with weights $p_j$ and
\begin{equation}\label{Tok}
    I_j=\frac{1}{R_j}(\varphi_j - \tilde{\varphi}).
\end{equation}

%in the same sense as $\tilde{n}(\omega,T_j)$ should be seen as mean occupancy of reservoirs.

Comparing Eq.~(\ref{Tok}) with Eq.~(\ref{J_el}), one can easily see that they have similar linear structure.
Indeed, for a particular $\kappa$ in Eq.~(\ref{J_el}), we can denote $1/R_j\rightarrow\omega_\kappa\gamma_{\kappa,j}$, $\varphi_j\rightarrow n_j(\omega_\kappa,T_j)$, $\tilde{\varphi}\rightarrow\tilde{n}(\omega_\kappa,\vec{T})$ in Eq.~(\ref{Tok}).
By that we instantly get ${J}_{\kappa,j}$ from $I_j$.
Thus, dissipation rates of OQS multiplied by modes' frequencies play the role of resistances at power minus one (conductivities), and occupancies of OQS's reservoirs are equivalent to potentials.

\begin{figure}
\center{\includegraphics[width=\linewidth]{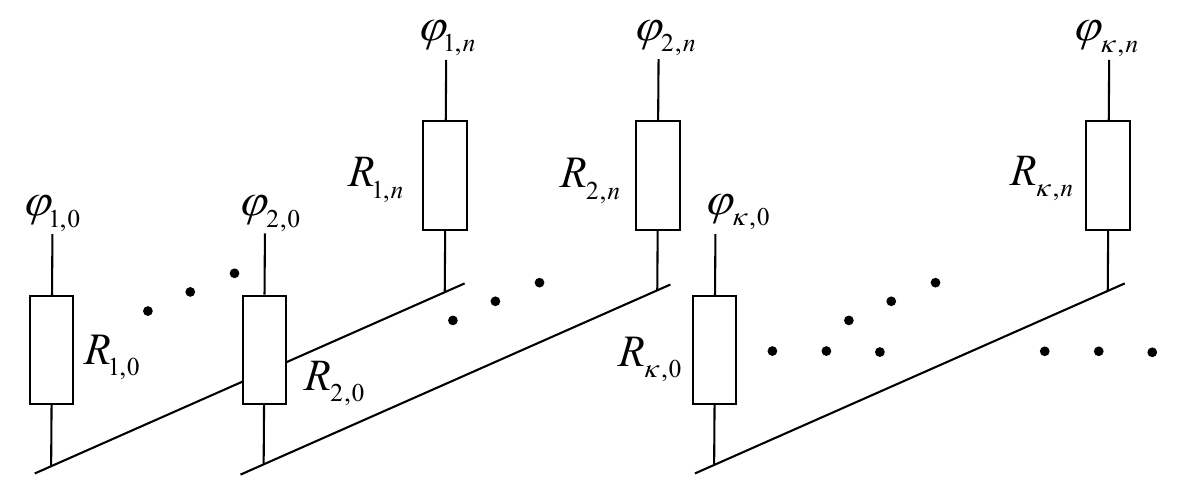}}
\caption{Electrical circuit supporting currents equivalent to the energy flows ${J}_{\kappa,j}(\vec{T})$ in Eq.~(\ref{J_el}).
This circuit is the circuit from Fig.~\ref{Scheme} repeated for each mode of the OQS.}
\label{Scheme1}
\end{figure}

\begin{figure}
\center{\includegraphics[width=\linewidth]{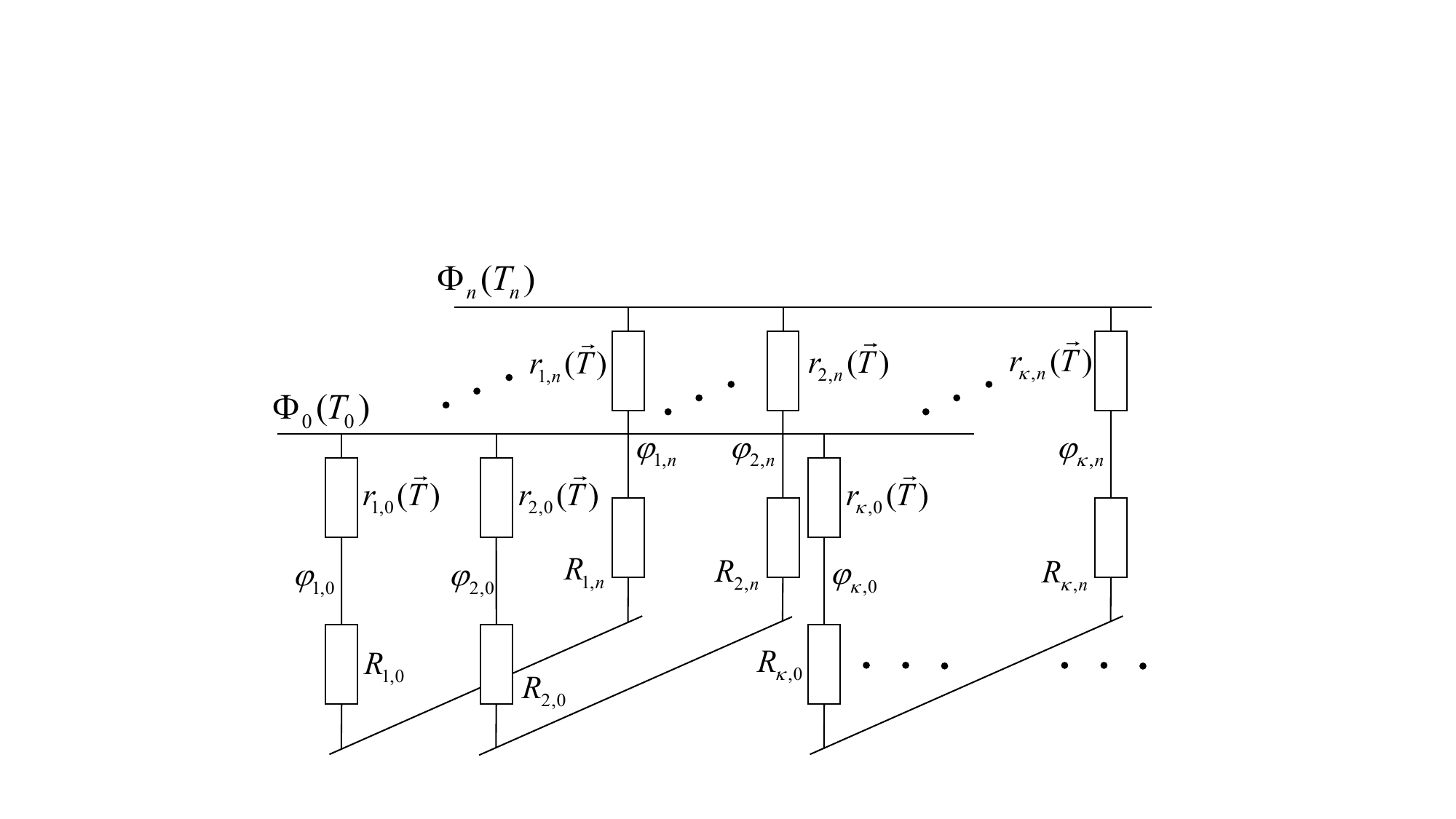}}
\caption{Equivalent CS of the OQS. Potentials $\varphi_{\kappa,j}$ from Fig.~\ref{Scheme1} 
are formed by the connection of the wires with the same $j$ by common bars through resistances $r_{\kappa,j}$.
The potentials $\Phi_j$ are applied to the bars.
The potentials $\Phi_j$ and the resistances $r_{\kappa,j}$ are defined by Eqs.~(\ref{PHI}) and provide the same currents through the wires as in Fig.~\ref{Scheme1}.}
\label{Scheme2}
\end{figure}

As there are many $\omega_\kappa$ modes in the OQS, the total system can be considered as a set of circuits from Fig.~\ref{Scheme}.
Transforming the central node into a bar with no resistance and repeating this circuit, we get the circuit in Fig.~\ref{Scheme1}.
Here we have denoted $R_{\kappa,j} \equiv 1/\omega_\kappa\gamma_{\kappa,j},\ 
    \varphi_{\kappa,j} \equiv n_j(\omega_\kappa,T_j)$ (note,  $\varphi_{\kappa,0}=0$).

If we connect all $\varphi_{\kappa,j}$ with the same $j$ through resistors to provide the decrease in potential, we get a crossbar representation of the considered OQS, see Fig.~\ref{Scheme2}.
Here
\begin{equation}\label{PHI}
\Phi_j(T_j) \equiv \varphi_{\kappa,j}+I_{\kappa,j}(\vec{T}) r_{\kappa,j}(\vec{T}),
\end{equation}
currents $I_{\kappa,j}(\vec{T})\equiv{J}_{\kappa,j}(\vec{T})/\omega_\kappa$.
Note, the $\Phi_j(T_j)$ is free from index $\kappa$ and depends only on $T_j$.

From Eqs.~(\ref{PHI}) it is seen that $\Phi_j(T_j)$ depend on $r_{\kappa,j}(\vec{T})$, while $r_{\kappa,j}(\vec{T})$ are free parameters.
This set of equations always has a solution.
Indeed, if potentials $\Phi_j(T_j)=\max_\kappa n_j(\omega_\kappa,T_j)$, then the resistances required for the decrease in potentials equal $r_{\kappa,j}(\vec{T})=(\Phi_j(T_j)-n_j(\omega_\kappa,T_j))/I_{\kappa,j}(\vec{T})$.
More over, this is not the unique solution of Eq.~(\ref{PHI}).
For $\Phi_j$ such that $\Phi_j>\max_\kappa n_j(\omega_\kappa,T_j)$, there always exists a set of resistors $r_{\kappa,j}$ that satisfies Eqs.~(\ref{PHI}).
Hence, there are infinitely many sets of $\Phi_j$ and $r_{\kappa,j}$ values that satisfy Eqs.~(\ref{PHI}).

For the considered computational scheme of vector-matrix multiplication based on OQS in each group of close modes, we have $\Phi_j(T_j)=n_j(\omega_\kappa/T_j)$ and $r_{\kappa,j}=0$.
%Also, $\varphi_0=0$.
%The particular choose from this set of solutions for $\Phi_j(T_j)$ allow to express temperature $T_j$ as function of chosen potential $\Phi_j$.
%Note that all potentials should be greater than zero, to be associated with some temperatures.

%Scheme in Fig.~\ref{Scheme2} is a crossbar structure by definition~\cite{de2021memristors,suri2020applications,liu2022memristor,adamatzky2013memristor}.
%Thus, we have succeeded in representing of the open quantum system as a crossbar structure.

As it is shown, electric currents in the circuit presented in Fig.~\ref{Scheme2} are equivalent to stationary energy flows in the OQS of bosonic modes connected to several reservoirs with different temperatures.
The dissipation rates in this OQS are equivalent to resistances at power minus one (conductivities), while reservoirs' occupancies and temperatures can be seen as potentials (see Figs.~\ref{Scheme}-\ref{Scheme2}).

Using this analogy, we can expand the class of matrices that can be used in computations similarly to how it is done in MNN.
If we need to multiply a matrix $A$ whose rows are normalized vectors with positive and negative elements by a positive vector $\vec{b}$, we can split it into two matrices $A=A_+-A_-$.
The matrix $A_+$ consists of all positive elements in their places in the matrix $A$ and zeros in other places.
The matrix $A_-$ consists of all absolute values of negative elements of $A$ in their places in $A$, and zeros in other places. 
After that, the computations can be done with both matrices $A_+$ and $A_-$ separately using Eq.~(\ref{vec_mat_mul}).
The results being subtracted with attenuation deliver $A\vec{b}$.

%vector $\vec{a}\in \mathds{R}^{n}$ has negative elements one can compute its scalar product with vector $\vec{b}\in \mathds{R}^{n}\ge 0$ separately for its positive and negative parts and subtract one from the other with needed attenuation.
%For that the procedure described above should be implemented separately for vector $\vec{a}_+$ that has same positive elements as $\vec{a}$ in the same positions and zeros in other positions and vector $\vec{a}_-=(\vec{a}_+-\vec{a})/2$ (vector $\vec{a}_-$ has zeros in the positions of positive elements of $\vec{a}$ and absolute values of negative elements of $\vec{a}$ in other positions).
%The second result should be subtracted from the first one.
%If the drain reservoir is regular radiation, this can be done by adding the energy flows with $\pi$ phase difference

%Note that, there is well known circuit approach to quantum transport \cite{pascal2011circuit,thomas2019photonic,dey2023negative,pekola2021colloquium}.
%However, it sufficiently differs from the electrical analogy developed in this paper.
%it is usually useful and transparent only in the same cases the Green's function approach does. %and it is hard to be applied to the system we consider.

%\section{Discussion and conclusion}
\textit{Conclusion and discussion.}
In this letter, we show that OQS of bosonic modes, i.e., photons, phonons, magnons, etc., can be used as {thermodynamic} coprocessors implementing multiple vector-matrix multiplications in a parallel analog way with stochastic matrices.
We show that stationary energy flow through one of the OQS's modes to a reservoir which is sufficiently colder than others (drain reservoir) 
%is proportional to the weighted occupancy of the other reservoirs, where weighted occupancy equals the sum of the reservoirs' occupancies and weighted dissipation rates products at the frequency of the oscillator.
%Thus, the stationary energy flow 
is proportional to the scalar product of vectors composed of reservoirs' occupancies and weighted dissipation rates.
We show that this fact and ability to manage dissipation rates of OQS's modes with close frequencies allow us to implement multiplication of an $n$-dimensional vector by an $m\times n$-dimensional matrix, where $n+1$ is the number of reservoirs and $m$ is the number of the modes having close frequencies.
The procedure requires the time needed for the establishment of the stationary energy flows, i.e., the OQS dissipation time, which does not depend on the
number of the reservoirs (input vector dimension) and the initial state of the OQS.
In parallel, vector-matrix multiplications with different matrices are implemented at other close frequency groups of OQS's modes.
%which might be useful for feature extraction and object recognition.
%However, the input vectors at these frequencies would be functionally dependent on the first one.

This computational system can be realized by microring resonators in the GHz range (OQS) and heated waveguides (reservoirs).
It is expected to reveal a maximal computational rate of $10\div1000$ TOps/s on the $5\times 5$ cm$^2$ area per a mode of the OQS.
However, with nowadays technological opportunities in temperature manipulation, this computational rate reduces to $100$ GOps/s.
Nevertheless, even this sufficiently reduced computational rate is comparable with the computational rate of some modern GPU solutions and is of interest in the context of a full-thermodynamic computer development.

The evolution of the OQS under interaction with the thermal environments is treated in the global approach to the dissipation~\cite{hofer2017markovian,potts2021thermodynamically,vovchenko2021model} that ensures the fulfillment of the second law of thermodynamics.
Hence, the computations being implemented by energy flows are accompanied by entropy growth due to OQS evolution towards its stationary state.
%The results of computations are produced by thermal environments.
%, i.e., the dissipation time of the OQS --- the time needed for the stationary energy flows through the OQS to be established.
%It depends on the dissipation rates of the OQS and barely depends on the number of entries (reservoirs).
Thus, the presented computational scheme can be seen as a part of developing fields of thermal linear algebra, thermodynamic computing, and thermal neural networks~\cite{aifer2024thermodynamic,melanson2025thermodynamic,lipka2024thermodynamic,tiwari2025quantum}.

Notably, intense interaction with the environment in many applications is considered to be a drawback, i.e., entanglement, coherence, and energy should be preserved as long as possible in the OQS ~\cite{scala2011robust,vovcenko2021dephasing,liu2022photonic}. 
In the considered computational scheme, the situation is opposite.
The faster OQS dissipates, the faster computations are done, particularly, the lower the Q-factor of the microring resonator is, the faster computations are done.

%we report on the existence of a crossbar structure inside OQS of quasi-particles with quadratic Hamiltonian (i.e. photons, phonons, magnons e.t.c.) coupled to several reservoirs with different temperatures.
%We show that by the construction of an electrical analogy for heat transport in this OQS.
%We demonstrate that stationary energy flows in the considered OQS can be mapped to electric currents, dissipation rates can be mapped to resistances at power minus one (conductivities), and reservoirs' occupancies and temperatures can be mapped to potentials in a circuit that can be represented as a crossbar structure.

We develop an electrical analogy for the discussed computing system.
We show that the system is equivalent to a CS.
The dissipation rates of the OQS multiplied by the OQS's modes' frequencies can be seen as conductivities, reservoirs' occupancies as potentials, and stationary energy flows as electric currents.
%used in MNN for vector-matrix multiplication.
%We discuss the limitations and possibilities of the learning process to be implemented on the found crossbar structure.
%We note that a similar analogy holds true for fermionic OQSs with quadratic Hamiltonians.
%Thus, the developed electrical analogy paves a way for usage of the OQSs as thermodynamics coprocessors.
%Also, it can be used to analyze transport properties of quantum devices via computational programs developed for classical electric circuits.

Also, we note that the similar dependence of energy and particle flows on the weighted occupancy of reservoirs is found in fermionic systems with quadratic Hamiltonians~\cite{vovchenko2025autonomous}, i.e., the same electrical analogy holds in this case.
Thus, these systems can also be used for the implementation of vector-matrix multiplication at the nanoscale.

%Finally, we underline some possible implementations of the considered system.
%The OQS used in the presented concept of an analog vector-matrix multiplier can be realized in plasmon lattice~\cite{hakala2018bose,kravets2018plasmonic}, waveguide array~\cite{yang2024programmable,pertsch1999optical} or cavity array~\cite{hartmann2006strongly,hartmann2008quantum,lee2010cooling}.
%In first realization plasmon particles can serve as reservoirs, while in second and third realizations reservoirs can be attached to waveguides and cavities.
%Propagating modes in these examples can be utilized for calculations.
%, being modes of the OQS.

%of the lattice can be seen as the modes of the OQS.
%In waveguide and cavity arrays, propagating modes can be seen as common modes of the open system, while reservoirs can be locally attached to waveguides and cavities.

%, for example, on the basis of interacting cavities, polarizes and spectrometers~\cite{}.
%Also note, that similar dependencies of energy and particles' flows on reservoirs' occupancies is found in fermionic systems with quadratic Hamiltinians~\cite{}.
%Thus, found crossbar structure can be also realized on the basis of nanoelectronic devices too.
%, and the experimental techniques of condensed matter physics can be used for modulation of dissipation rates

%Note that, there is well known circuit approach to quantum transport \cite{pascal2011circuit,thomas2019photonic,dey2023negative,pekola2021colloquium}.
%However, it sufficiently differs from the electrical analogy developed in this paper.

\textbf{Acknowledgement.}
%This work was financially supported by the Russian Science Foundation (project no. 20-72-10057).
A.A.Z. and E.S.A. acknowledge the support of the Foundation for the Advancement of Theoretical Physics and Mathematics BASIS.

\bibliography{Cross_Bar}

\end{document}